# Random Matrix Theory and Cross-correlations in Global Financial Indices and Local Stock Market Indices

Ashadun Nobi

*Department of Physics, Inha University, Incheon 402-751, Korea, and*

*Department of Computer Science and Telecommunication Engineering, Noakhali Science*

*and Technology University, Sonapur Noakhali-3802, Bangladesh*

Seong Eun Maeng, Gyeong Gyun Ha, and Jae Woo Lee

*Department of Physics, Inha University, Incheon 402-751, Korea*

We analyzed cross-correlations between price fluctuations of global financial indices (20 daily stock indices over the world) and local indices (daily indices of 200 companies in the Korean stock market) by using random matrix theory (RMT). We compared eigenvalues and components of the largest and the second largest eigenvectors of the cross-correlation matrix before, during, and after the global financial the crisis in the year 2008. We find that the majority of its eigenvalues fall within the RMT bounds [$\lambda_-$, $\lambda_+$], where $\lambda_-$ and $\lambda_+$ are the lower and the upper bounds of the eigenvalues of random correlation matrices. The components of the eigenvectors for the largest positive eigenvalues indicate the identical financial market mode dominating the global and local indices. On the other hand, the components of the eigenvector corresponding to the second largest eigenvalue are positive and negative values alternatively. The components before the crisis change sign during the crisis, and those during the crisis change sign after the crisis. The largest inverse participation ratio (IPR) corresponding to the smallest eigenvector is higher after the crisis than during any other periods in the global and local indices. During the global financial the crisis, the correlations among the global indices and among the local stock indices are perturbed significantly. However, the correlations between indices quickly recover the trends before the crisis.

Email: jaewlee@inha.ac.kr

Fax: +82-32-872-7562

# I. INTRODUCTION

In recent years, statistical characterizations of financial markets based on the concepts and the methods of physics have attracted considerable attention [1, 2]. Different models and theoretical approaches have been developed to describe the features of financial dynamics [3-6]. Recently, the correlation and the network properties of 20 global financial indices have been investigated successfully [7]. Our main focus is to understand the global financial structure and the local stock market after a crisis. We investigated the recovery and the reformation of the global financial indices and the local stock market after a crisis and its comparison with those before and during the crisis. Conventionally, trade protectionism increases during a crisis. We think that if the market situation after a crisis is understood better, more investment can be made in the market, which can push the economy up.

Random matrix theory (RMT) has been applied to investigate the statistical properties of the cross-correlations of price changes of global financial indices and the local stock market. It was developed in the context of complex quantum systems in which the precise nature of the interactions between subunits is not known [8, 9]. For complex quantum systems, RMT predictions represent an average over all possible interactions [10-12]. Deviations from RMT predictions identify non-random properties of the system under consideration, providing clues about the underlying interactions [8, 13]. The eigenvalues of the cross-correlation matrix are compared with the eigenvalues of the random matrix, and we find that some eigenvalues deviate from the RMT prediction. The largest eigenvalue and its corresponding eigenvector components represent the collective 'response' of the entire market to stimuli. We find that the largest eigenvalue during a crisis is higher than that in any other period, which indicates strong interactions among indices during a crisis. The components of the eigenvectors corresponding to the largest eigenvalues in all periods are positive, representing an influence that is common to all financial indices. The components of the eigenvectors corresponding to the second largest eigenvalue are positive and negative alternatively, and during a crisis some of them change their sign. The eigenvectors corresponding to the remaining eigenvalues are in the range of the RMT predictions, and they do not show any significant difference due to a crisis.

This paper is organized as follows: Section II contains a brief description of the data set analyzed in this work. Section III discusses the eigenvalue distribution of the cross-correlation matrix, and its comparison with the RMT. Section IV includes the analysis results for the cross-correlation matrix. Finally, Section V contains some concluding remarks.

## II. DATA ANALYZED

We analyze the daily closing prices (global indices) of 20 financial markets of the world and the daily closing prices (local indices) of 200 Korean stocks from the period June 2, 2006, to July 29, 2011. Global financial indices are as follows: Argentina, MERV; Brazil, BVSP; Egypt, CCSI; India, BESAN; Indonesia, JKSE; Malaysia, KLSE; Mexico, MXX; South Korea, KS11; Taiwan, TWII; Australia, AORD; Austria, ATX; France, FCHI; Germany, GDAXI; Hong Kong, HIS; Israel, TA100; Japan, N225; Singapore, STI; Switzerland, SSMI; the United Kingdom, FTSE; and the United States, GSPC. The data are collected from Ref. 14. Local indices consist of the daily indices of 200 companies belonging to the Korean stock market. We divided the data into three periods. We considered the period from June 2, 2006, to November 30, 2007, as before the crisis, from December 3, 2007, to June 30, 2009, as during the crisis, and from January 1, 2010, to June 30, 2011, as after the crisis. During the crisis period, Lehman Brothers declared bankruptcy on November 15, 2008. This collapse ignited the global financial the crisis. To make an equal-time cross-correlation matrix, we remove some days on the basis of public holidays. If 30% of the markets are not open on a specific day, we remove that day. Again, we add some days for a specific market if that market is closed on a particular day. In this case, we consider the last closing prices for that day. Thus, we considered all indices at the same date and filtered the data as in Ref. 15. We examined daily returns for the indices, each containing 388 records. The financial crisis of 2007-2009 is known to be the worst financial the crisis since the Great Depression of the 1930s; and it originated in the United States and then spread to the world [16].

## III. RANDOM MATRIX THEORY APPROACH

Let $P_i(t)$ be the daily closing prices at time t of indices $i = 1, \cdots, N$ where N is the total number of indices. The time spans $t = 1, \cdots, T$, where T is the maximum time of each window. The change of price at time t is given by

$$R_i(t) = \ln P_i(t+1) - \ln P_i(t). \tag{1}$$

Because different stocks have varying levels of volatility (standard deviation), we define a normalized return:

$$r_i(t) = \frac{R_i(t) - <R_i(t)>}{\sigma_i}, \tag{2}$$

where $\sigma_i$ is the standard deviation of $R_i$, and $< \cdots >$ represents the time average.

The cross-correlation matrix C is expressed in terms of $r_i(t)$ as

$$C_{ij} = <r_i(t)r_j(t)>, \tag{3}$$

where C is a real, symmetric matrix with $C_{ii} = 1$ and $C_{ij}$ has values in the range [-1,1]. Then, we compare the properties of C with those of a random cross-correlation matrix (Wishart matrix) [17, 18]. The statistical properties of random matrices are known [19]. Especially, in the limit $N \to \infty, L \to \infty$ with $Q = L/N (\geq 1)$ for N time series and L random elements with zero mean and unit variance, the probability density function $P_{rm}(\lambda)$ of the eigenvalues λ of the random correlation matrix is given by

$$P_{rm}(\lambda) = \frac{Q}{2\pi} \frac{\sqrt{(\lambda_+ - \lambda)(\lambda - \lambda_-)}}{\lambda} \tag{4}$$

for λ within the bounds $\lambda_- \leq \lambda \leq \lambda_+$, where $\lambda_-$ and $\lambda_+$ are the minimum and the maximum eigenvalues of random matrix, respectively, given by

$$\lambda_\pm = 1 + \frac{1}{Q} \pm 2\sqrt{\frac{1}{Q}}. \tag{5}$$

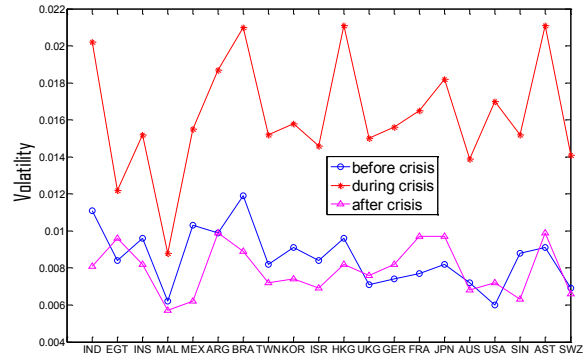

(a)

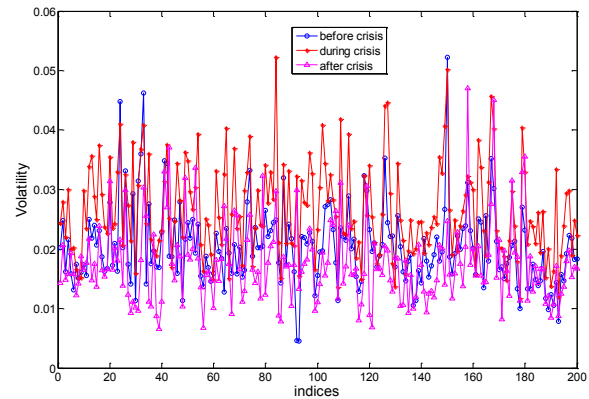

(b)

Fig. 1. Volatility (standard deviation) as a measure of fluctuations (a) in global financial indices and (b) in the local indices (Korean stock market indices) before, during, and after the 2008 financial the crisis. In the global market, the volatilities were enhanced during the crisis period. In the local market, the volatilities increased slightly during the crisis.

## IV. GLOBAL AND LOCAL INDICES

We measured the effects of the global financial crisis on the global stock market and the local stock market. The financial the crisis influenced the global market greatly. Figure 1 showed the volatilities (standard deviations) of three time windows for the world market and the local Korean stock market. The volatilities of the world indices increase almost two times during the crisis period. However, in the local Korean stock market, the volatilities for individual companies increase only slightly during the crisis. Of course, the net effects of the volatilities on the whole stock market in Korea were great, as we observe in Fig. 1(a).

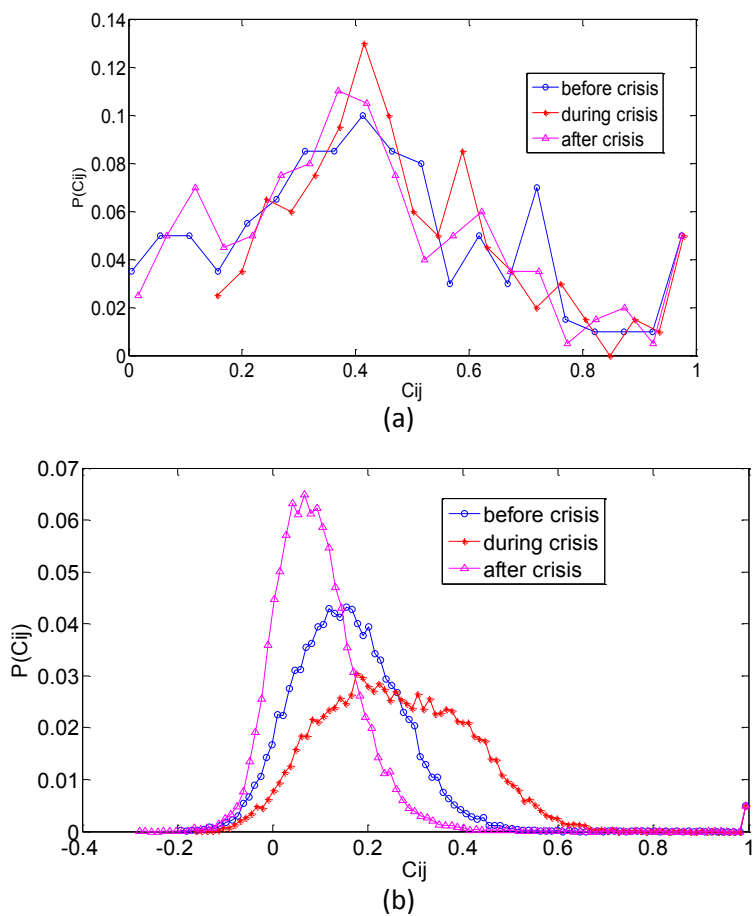

FIG. 2. Probability density of the cross-correlation matrix $C_{ij}$ calculated by using daily returns of (a) 20 world stock indices and (b) 200 stock prices of individual companies in the Korean stock market before, during and after the 2008 financial crisis.

We calculated the cross-correlation matrix of the price changes in the global and the local markets. In Figure 2, we presented the probability distributions for the components of cross-correlation matrix for the world market and the Korean stock market. The average values of the cross-correlation coefficients are 0.4283 before the crisis, 0.4859 during the crisis, and 0.4197 after the crisis in the global stock indices. The average cross-correlation coefficient during the crisis period is higher than those before and after the crisis. The standard deviations of the cross-correlation coefficients are 0.2353 before the crisis, 0.1966 during the crisis and 0.2235 after the crisis in the global stock indices. During the crisis period the distribution of the cross-correlation coefficients is narrower, and the standard deviation decreases. Almost similar behaviors are observed in the local stock market. In the Korean stock market, we observed average values: 0.1613 before the crisis, 0.2609 during the crisis and 0.0947 after the crisis. We also calculated the standard deviations: 0.1076 before the crisis, 0.1249 during the crisis, and 0.1002 after the crisis. In Korean stock market, the average cross-correlation coefficient during the crisis is higher than that of before and after the crisis, and the cross-correlation coefficients distributed broadly during the crisis period. After the crisis the distribution becomes narrower.

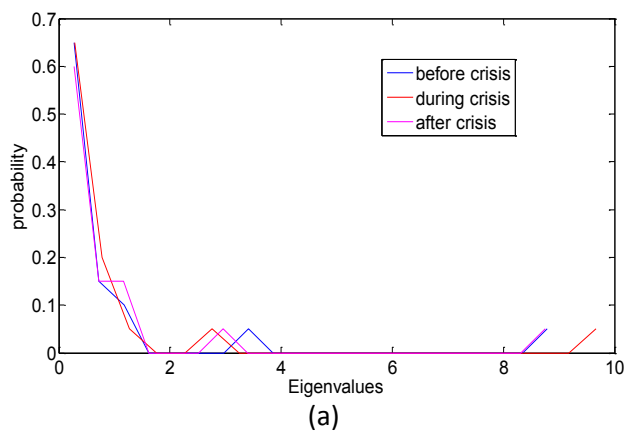

(a)

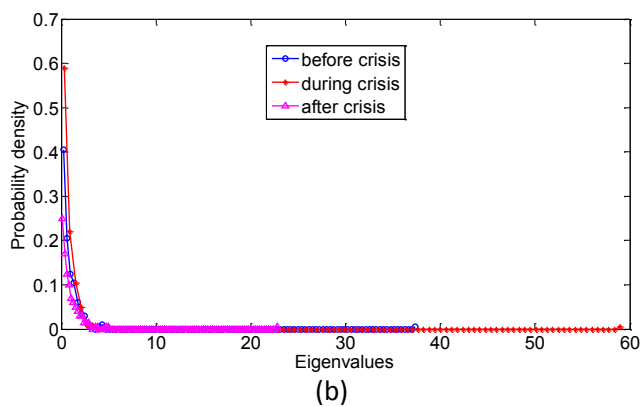

(b)
6

Fig. 3. Probability density of eigenvalues calculated by (a) using daily return of 20 world financial indices and (b) 200 local Korean stock prices of individual companies before, during and after the 2008 financial crisis.

The cross-correlation matrix of strongly correlated stock indices shows structure very different from that of a random matrix theory. For a real market some eigenvalues deviate from RMT predictions, which has been confirmed by several studies [20-22]. In random matrix theory, the eigenvalues are bounded on the range $\lambda_- \leq \lambda \leq \lambda_+$, where the lower and the upper bounds of the eigenvalue are given by Eq. (5). In the global stock indices, we obtain $\lambda_+$=1.506 and $\lambda_-$=0.597 by inserting the factor Q=19.4 into Eq.(5). This indicates that for an uncorrelated time series, the eigenvalues should be bounded between 0.597 and 1.506. We compute the eigenvalues of the empirical matrix $C$ for all periods, and we find that some eigenvalues deviate from RMT predictions. We obtained the maximum (minimum) eigenvalues as $\lambda_{min(max)}$=0.0521 (9.0143) before the crisis, $\lambda_{min(max)}$=0.0392 (9.9189) during the crisis, and $\lambda_{min(max)}$=0.0549 (8.9818) after the crisis. The larger eigenvalue during the crisis shows that there is a stronger correlation among financial indices during the crisis. The largest eigenvalue after the crisis is lower, which indicate that the crisis has been overcome and the correlation among indices is comparatively less strong than other periods. The differences in the probability densities of the cross-correlation coefficients and the eigenvalue between before and after the crisis indicate that after the crisis, the market becomes more stable, with less interaction among the indices than before the crisis. The better stability of the Korean market after the crisis was observed using the volatility. Figure 3 compares the probability distributions of eigenvalues for all periods. In the local Korean stock market, we examine daily returns for N=200 indices, each containing L=390. Thus, we has Q=1.95, and we obtain $\lambda_+$=2.945 and $\lambda_-$=0.08. The eigenvalues of the empirical matrix $C$ for all periods also deviate from RMT predictions in the local market. The maximum (minimum) eigenvalues are $\lambda_{min(max)}$=0.02078 (37.56) before the crisis, $\lambda_{min(max)}$=0.0207 (59.26) during the crisis, and $\lambda_{min(max)}$=0.0223 (22.90) after the crisis. The local market is also correlated strongly during the crisis and recovers to the ordinary state after the crisis.

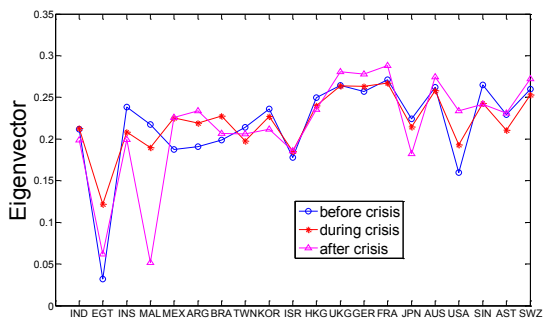

(a)

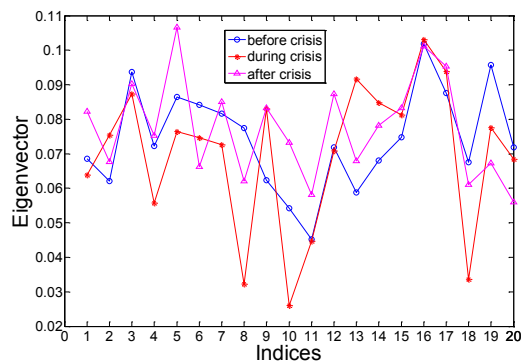

(c)

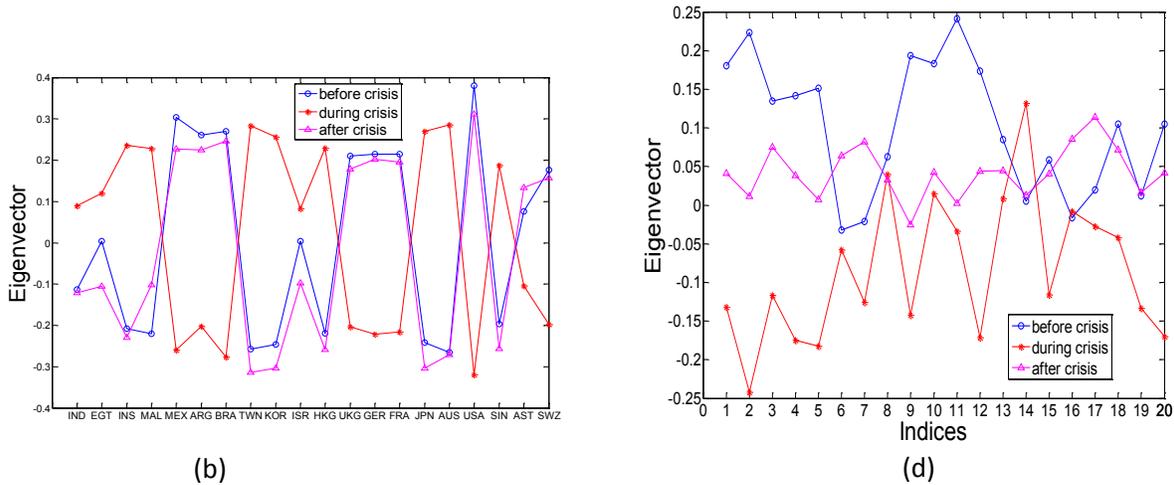

Fig. 4. Comparison of the components for (a) the largest and (b) the second largest eigenvectors of 20 financial indices before, during and after the 2008 financial crisis. The components for (c) the largest and (d) the second largest eigenvectors in the local stock indices before, during, and after the 2008 financial crisis are plotted. Only the 20 largest companies are plotted because the figure becomes crowded if all components are plotted.

The eigenvectors corresponding to the largest eigenvalue are shown in Fig. 4(a) for the global market and in Fig. 4(c) for the local Korean stock market. The components of the eigenvectors corresponding to the second largest eigenvalue are shown in Fig. 4(b) and (d). In the global market and the local Korean stock market, all the components carry the same sign which represents the same market mode, and there is no significant difference due to the crisis. However, after the crisis, a significant change is seen for the component of Malaysia. In the global market, we find that the components of an eigenvector that carries a negative sign (such as Argentina, Brazil, etc.) during the crisis change their sign after the crisis and that the components of an eigenvector that bears a positive sign (India, Egypt, etc.) during the crisis show opposite signs after the crisis. We also find the same kind of behavior between the components of the eigenvector before and during the crisis [7]. The eigenvectors corresponding to the eigenvalues near or between RMT predictions do not show any significant behavior in the global market indices. We find that most of the components of the second largest eigenvector in the local Korean stock market switch to the opposite directions during the crisis. Generally, the indices which show large volatility (during the crisis) move opposite direction during the crisis. The eigenvectors corresponding to the eigenvalues near or between RMT predictions do not show any significant behavior.

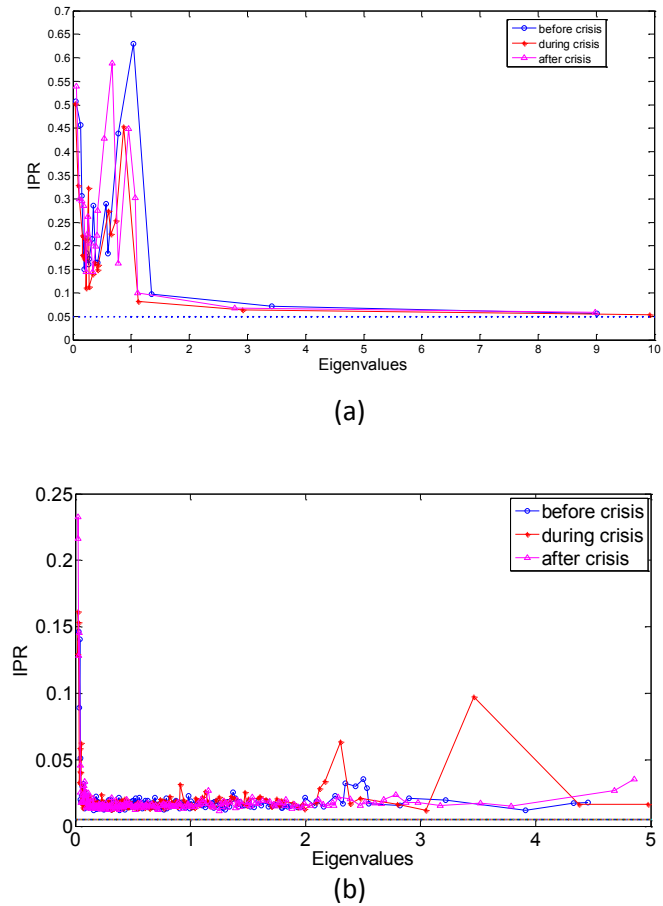

Fig. 5. Comparison of the inverse participation ratio for (a) the global financial indices and (b) the local Korean stock indices before (circle), during (plus) and after the 2008 financial crisis (triangles). There is no significance difference between the periods. The dotted line in the global market marks the value 0.05 of the IPR. The dotted line in the local Korean stock market marks the value 0.005 of the IPR when all components contribute equally.

The inverse participation ratio (IPR) of the eigenvector $u^k$ is defined as $I^k = \sum_{l=1}^{N}[u_l^k]$, where $u_l^k$, $l = 1, \cdots, N$, are the components of eigenvector $u^k$. A comparison of the IPR is shown in Fig. 5. The IPR quantifies the reciprocal of the number of eigenvector components that contribute significantly [21, 22]. In the global market, no significant difference is seen in any period. In the local Korean stock market, the largest eigenvalue is $1/I^k \approx 148$ during the crisis, 140 before the crisis, and 119 after the crisis. This indicates that, during the crisis, more stocks participate in the largest eigenvector. In addition, we observe that the largest IPR is 0.2328 after the crisis, 0.1614 during the crisis, and 0.1462 before the crisis, which indicates that comparatively few stocks participate on the smallest eigenvector after the crisis.

## V. CONCLUSION

We analyzed the cross-correlation matrices of stock price changes in global financial indices and local Korean stock indices before, during and after the 2008 financial crisis. We calculated the eigenvalues and eigenvectors of the correlation matrix. Almost all eigenvalues were in the predicted range of random matrix theory. However, some eigenvalues deviated from the predictions of random matrix theory for the global indices and the local Korean indices. We observed that eigenvalues during the crisis were higher than they were during other periods. Then, we investigated the components of the largest and the second largest eigenvectors. We observed that the components of the second largest eigenvectors for the global and the local indices showed the same behaviors before and after the crisis but during the crisis, they showed opposite behaviors. We also observed that the largest inverse participation ratio (IPR) corresponding to the smallest eigenvector in the global and local stock indices after the crisis was higher than it was during other periods.

## ACKNOWLEDGEMENT

This research has been supported by the research fund of Inha University.

## REFERENCES

[1] R. N. Mantegna and H. E. Stanely, *An introduction to Econophysics* (Cambridge University Press, Cambridge, 2000).

[2] J. P. Bouchaud and M. Potters, *Theory of Financial Risk* (Cambridge University Press, Cambridge , 2000).

[3] D. Challet and Y. C. Zhang, Physica A **246**, 407 (1997).

[4] T. Lux and M. Marchesi, Nature **397**, 498 (1999).

[5] P. Stauffer, P. M. C De Oliveria and A. T. Bernardes, Int. J. Theor. Appl. Finance. **2**, 83 (1999).

[6] V. M. Eguiluz and M. G. Zimmermann, Phys. Rev. Lett. **85**, 5659 (2000).

[7] S. Kumar and N. Deo, Phys. Rev. E **86**, 026101 (2012).

[8] M. L. Mehta, *Random Matrices* (Academic Press, Boston, 1991).

[9] T. Guhr, A. Müller-Groeling, and H. A. Weidenmüller, Phys. Rep. **299**, 190 (1998).

[10] F. J. Dyson and M. L. Mehta, J. Math. Phys. **4**, 701 (1963).

[11] F. J. Dyson, J. Math. Phys. **3**, 140 (1962).

[12] M. L. Mehta and F. J. Dyson, J. Math. Phys. **4**, 713 (1963).

[13] T. A. Brody, J. Flores, J. B. French, P. A. Mello, A. Pandey and S. S. M. Wong, Rev. Mod. Phys. **53**, 385 (1981).

[14] http://finance.yahoo.com.

[15] R. K. Pan and S. Sinha, Phys. Rev. E **76**, 046116 (2007); L. Sandoval, Jr. and I. D. P Franca, Physica A **391**, 187 (2012).

[16] D-M. Song, M. Tumminello, W-X. Zhou and R. N. Mantegna, Phys. Rev. E **84**, 026108 (2011).

[17] T. H. Baker, P. J. Forrester and P. A. Pearce, J. Phys. A **31**, 6087 (1998).

[18] A. Edelman, SIAM J. Matrix Anal. Appl. **9**, 543 (1998).

[19] A. M. Sengupta and P. P. Mitra, Phys. Rev. E **60**, 3389 (1999).

[20] L. Laloux, P. Cizeau, J-P. Bouchaud and M. Potters, Phys. Rev. Lett. **83**, 1467 (1999).

[21] V. Plerou, P. Gopikrishnan, B. Rosenow, L. A. N. Amaral and H. E. Stanley, Phys. Rev. Lett. **83**, 1471 (1999).

[22] V. Plerou, P. Gopikrishnan, B. Rosenow, L. A. N. Amaral, T. Guhr and H. E. Staneley, Phys. Rev. E **65**, 066126 (2002).